\documentclass[11pt]{article}
\usepackage{graphicx}
\usepackage{dcolumn}
\usepackage{amsmath}
\usepackage{epsfig}
\RequirePackage{xspace}
\def\CP     {\ensuremath{C\!P}\xspace}
\def\Bbar   {\kern 0.18em\overline{\kern -0.18em B}{}\xspace}
\def\Dbar   {\kern 0.2em\overline{\kern -0.2em D}{}\xspace}
\def\Db     {\ensuremath{\Dbar}\xspace}
\def\jpsi   {\ensuremath{{J\mskip -3mu/\mskip -2mu\psi\mskip 2mu}}\xspace}
\def\psitwos{\ensuremath{\psi{(2S)}}\xspace}
\def\KS     {\ensuremath{K^0_{\scriptscriptstyle S}}\xspace} 
\def\KL     {\ensuremath{K^0_{\scriptscriptstyle L}}\xspace} 
\mathchardef\Upsilon="7107
\def\Y#1S{\ensuremath{\Upsilon{(#1S)}}\xspace}
\def\FourS  {\Y4S}
\def\nub    {\ensuremath{\overline{\nu}}\xspace}
\def\nutb   {\ensuremath{\nub_\tau}\xspace}
\def\nutb   {\ensuremath{\nub_\tau}\xspace}
\newcommand {\gevcc}{\ensuremath{{\mathrm{\,Ge\kern -0.1em V\!/}c^2}}\xspace}
\newcommand {\tevcc}{\ensuremath{{\mathrm{\,Te\kern -0.1em V\!/}c^2}}\xspace}
\def\invfb  {\ensuremath{\mbox{\,fb}^{-1}}\xspace}

\begin{document}

\begin{center}
{\Large\bf
{\boldmath $\CP$} violation and hints for new physics at the {\boldmath $B$} factories}
\end{center}
\bigskip

\begin{center}
\large
  Gagan B. Mohanty\\
  (on behalf of the Belle and BaBar Collaborations)\\
  {\em Tata Institute of Fundamental Research,}\\
  {\em Homi Bhabha Road, Mumbai 400 005, India}
\end{center}
\bigskip\bigskip

\begin{abstract}
We report the latest results on \CP\ violation measurements and the
tantalizing hints of potential new physics effects obtained at the $B$
factories.
\end{abstract}

\section{Introduction}
\label{intro}

In the standard model (SM) of particle physics, \CP\ violation occurs
due to an irreducible phase appearing in the quark-flavor mixing matrix,
called the Cabibbo-Kobayashi-Maskawa (CKM) matrix, which relates the weak
interaction eigenstates to that of mass. The study of $B$ meson decays
allows us to carry out a multitude of measurements involving
the angles and sides of the so-called unitarity triangle (UT), a graphical
sketch of the unitarity of the CKM matrix in the complex plane. The
{\em raison d'\^{e}tre} of the two $B$-factory experiments -- Belle at
KEK, Japan and BaBar at SLAC, USA -- was to precisely measure various UT
parameters. By doing so, they were designed to verify the \CP\ violation
mechanism within the SM, as suggested by Kobayashi and Maskawa~\cite{km},
and to set constraints on potential new physics contributions in the flavor
sector.

In these proceedings, we summarize recent results on \CP\ violation,
involving three angles of the unitarity triangle, and describe a number
of hints for new physics observed with the $B$-factory experiments.

\section{Angles of the unitarity triangle}
\label{ang}

The UT angles are determined through the measurement of the time dependent
\CP\ asymmetry, $A_{\CP}(t)$, defined as
\begin{equation}
A_{\CP}(t) = \frac{N[\Bbar^0(t)\to f_{\CP}]-N[B^0(t)\to f_{\CP}]}
                  {N[\Bbar^0(t)\to f_{\CP}]+N[B^0(t)\to f_{\CP}]},
\label{eq1}
\end{equation}
where $N[\Bbar^0/B^0(t)\to f_{\CP}]$ is the number of $\Bbar^0/B^0$s that
decay into a \CP\ eigenstate $f_{\CP}$ after time $t$. The asymmetry, in
general, can be expressed in terms of two components:
\begin{equation}
A_{\CP}(t)=S_f\sin(\Delta mt)+A_f\cos(\Delta mt),
\label{eq2}
\end{equation}
where $\Delta m$ is the difference in mass of $B^0$ mass eigenstates.
The sine coefficient $S_f$ is related to the UT angles, while the cosine
coefficient $A_f$ is a measure of direct \CP\ violation. For the latter
to have a nonzero value, we need at least two competing amplitudes with
different weak and strong phase to contribute to the decay final state.
As an example, for the decay $B^0\to\jpsi\KS$, where mostly one diagram
contributes, the cosine term is expected to vanish and the sine term is
proportional to the UT angle $\phi_1$\footnote{An alternative notation
of $\beta$, $\alpha$ and $\gamma$, that correspond to $\phi_1$, $\phi_2$
and $\phi_3$, respectively, is equally abundant in the literature.}. The
time-dependent \CP\ asymmetry is, therefore, given as
\begin{equation}
A_{\CP}(t) = -\xi_f\sin(2\phi_1)\sin(\Delta mt),
\label{eq3}
\end{equation}
where $\xi_f$ is the \CP\ eigenvalue of the final state. In the case of
$B$ factories, the measurement of $A_{\CP}(t)$ utilizes decays of the
\FourS\ into two neutral $B$ mesons, of which one can be completely
reconstructed into a \CP\ eigenstate, while the decay products of the
other (called the tag $B$) identify its flavor at decay time. The time
difference $t$ between the two $B$ decays is determined by reconstructing
their decay vertices. Finally the \CP\ asymmetry amplitudes, proportional
to the UT angles, are obtained from an unbinned maximum likelihood fit to
the proper time distributions separately for events tagged as $\Bbar^0$
and $B^0$.

\subsection{The angle {\boldmath $\phi_1$}}

\begin{figure}[!htb]
\center
\includegraphics[width=.7\columnwidth]{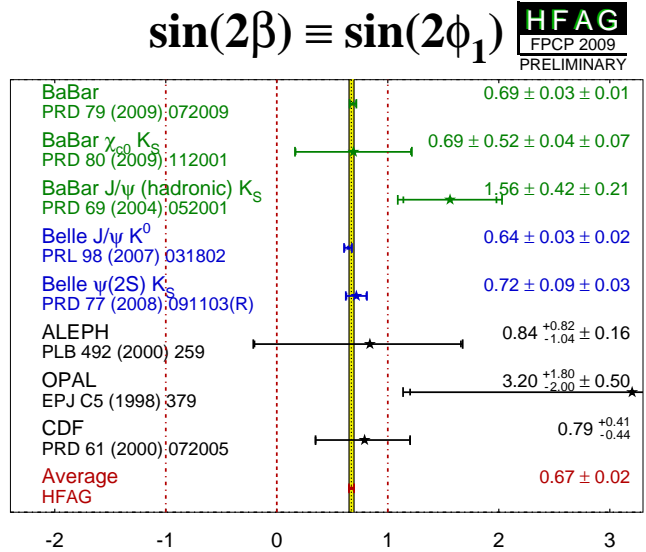}
\caption{Average of $\sin(2\phi_1)$ from all experiments, as compiled by the
         HFAG~\cite{hfag}.}
\label{sin2phi1}
\end{figure}

The most precise measurement of the angle $\phi_1$ is obtained from a
study of the decays $B^0\to$ charmonium $+K^{(*)0}$. These decays, known
as ``golden modes'', mainly proceed via the CKM-favored tree diagram
$b\to c\bar{c}s$ with an internal $W$ boson emission. The subleading
penguin (loop) contribution to the final state, that has a different weak
phase compared to the tree diagram, is suppressed by almost two orders of
magnitude. This makes $A_f=0$ in Eq.~\ref{eq2} to a very good approximation.
Besides the theoretical simplicity, these channels also offer experimental
advantages because of the relatively large branching fractions ($\sim 10^{-3}$)
and the presence of narrow resonances in the final state, which provides
a powerful rejection against combinatorial background. The \CP\ eigenstates
considered for this analysis include $\jpsi\KS$, $\psitwos\KS$, $\chi_{c0}\KS$,
$\eta_c\KS$ and $\jpsi\KL$. The measured world-average value of $\sin(2\phi_1)$
is $0.67\pm 0.02$. Figure~\ref{sin2phi1} shows the impact of this measurement
by Belle and BaBar, that eventually led to half of the 2008 physics Nobel
prize~\cite{nobel} being awarded to Kobayashi and Maskawa, when compared to
other experiments.

\subsection{The angle {\boldmath $\phi_2$}}

Decays of $B$ mesons to the final states $hh$ ($h=\rho$ or $\pi$),
dominated by the CKM-suppressed $b\to u$ transition, are sensitive
to the UT angle $\phi_2$. The presence of $b\to d$ penguin diagrams,
however, complicates the situation by introducing additional phases
such that the measured parameter is no more $\phi_2$ alone, rather
an effective value $\phi^{\rm eff}_2 =\phi_2+\delta\phi_2$. (Note that
the same prescription {\it vis-a-vis} penguin pollution also applies
to other UT angles, wherever appropriate.) At present, the most
precise measurement of this angle is obtained in the analysis of
the decays $B\to\rho\rho$. Combining with additional constraints
coming from $B\to\rho\pi$ and $B\to\pi\pi$, we measure $\phi_2=
\left(89.0^{+4.4}_{-4.2}\right)^\circ$~\cite{ckmfitter}.

\subsection{The angle {\boldmath $\phi_3$}}

The angle $\phi_3$ is measured by exploiting the interference between the
decays $B^-\to D^{(*)0}K^{(*)-}$ and $B^-\to\Db^{(*)0}K^{(*)-}$, where
both $D^0$ and $\Dbar^0$ decay to a common final state. This measurement
can be performed in three different ways: utilizing decays of $D$ mesons
to \CP\ eigenstates~\cite{glw}, making use of doubly Cabibbo-suppressed
decays of the $D$ meson~\cite{ads}, and exploiting the interference
pattern in the Dalitz plot of $D\to\KS\pi^+\pi^-$ decays~\cite{ggsz}.
Currently, the last method provides the strongest constraint on $\phi_3$.
Combining all related measurements from Belle and BaBar, the world-average
value is found to be $\phi_3=\left(73^{+19}_{-24}\right)^\circ$~\cite{ckmfitter}.

\begin{figure}[!htb]
\center
\includegraphics[width=.48\columnwidth]{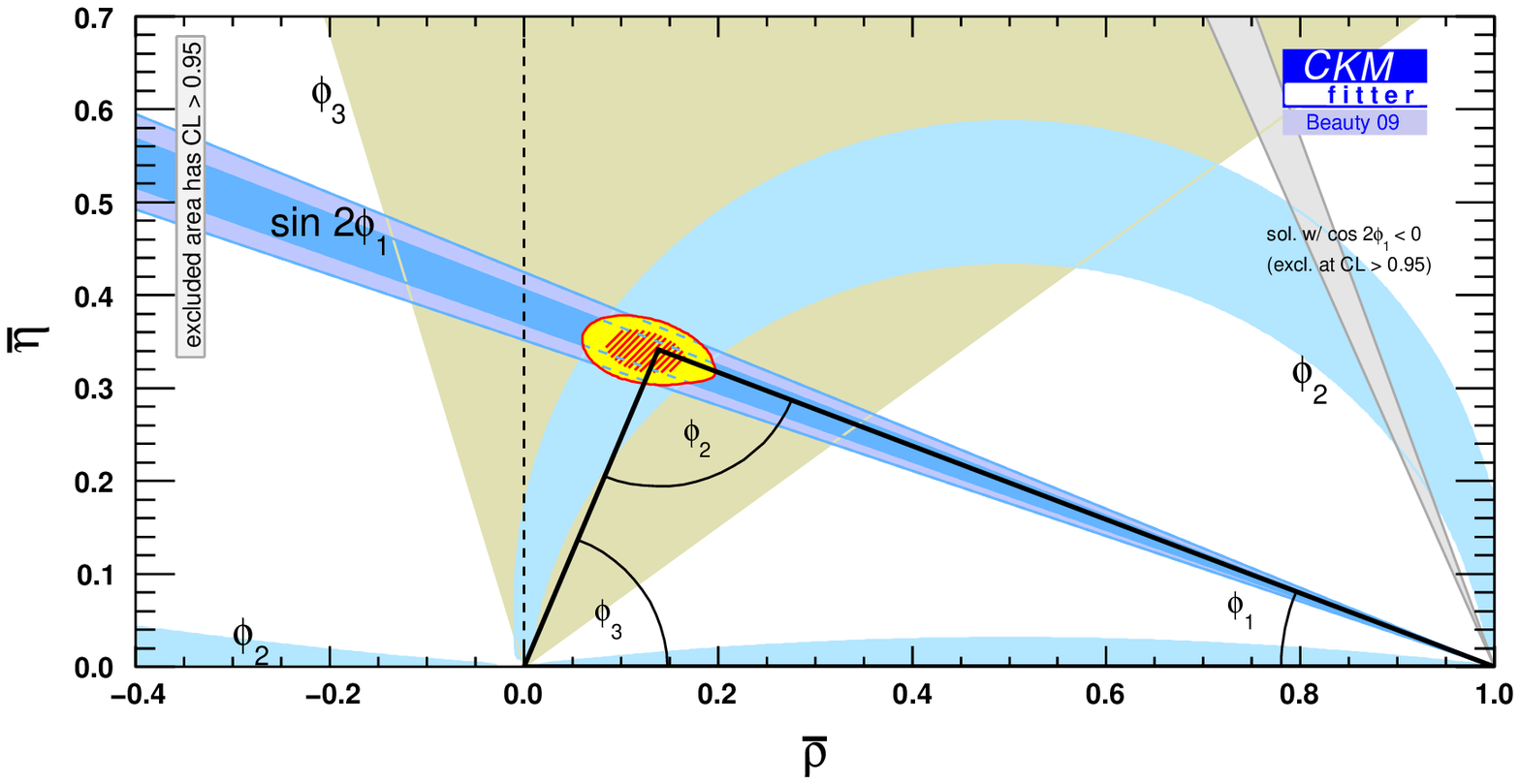}
\includegraphics[width=.48\columnwidth]{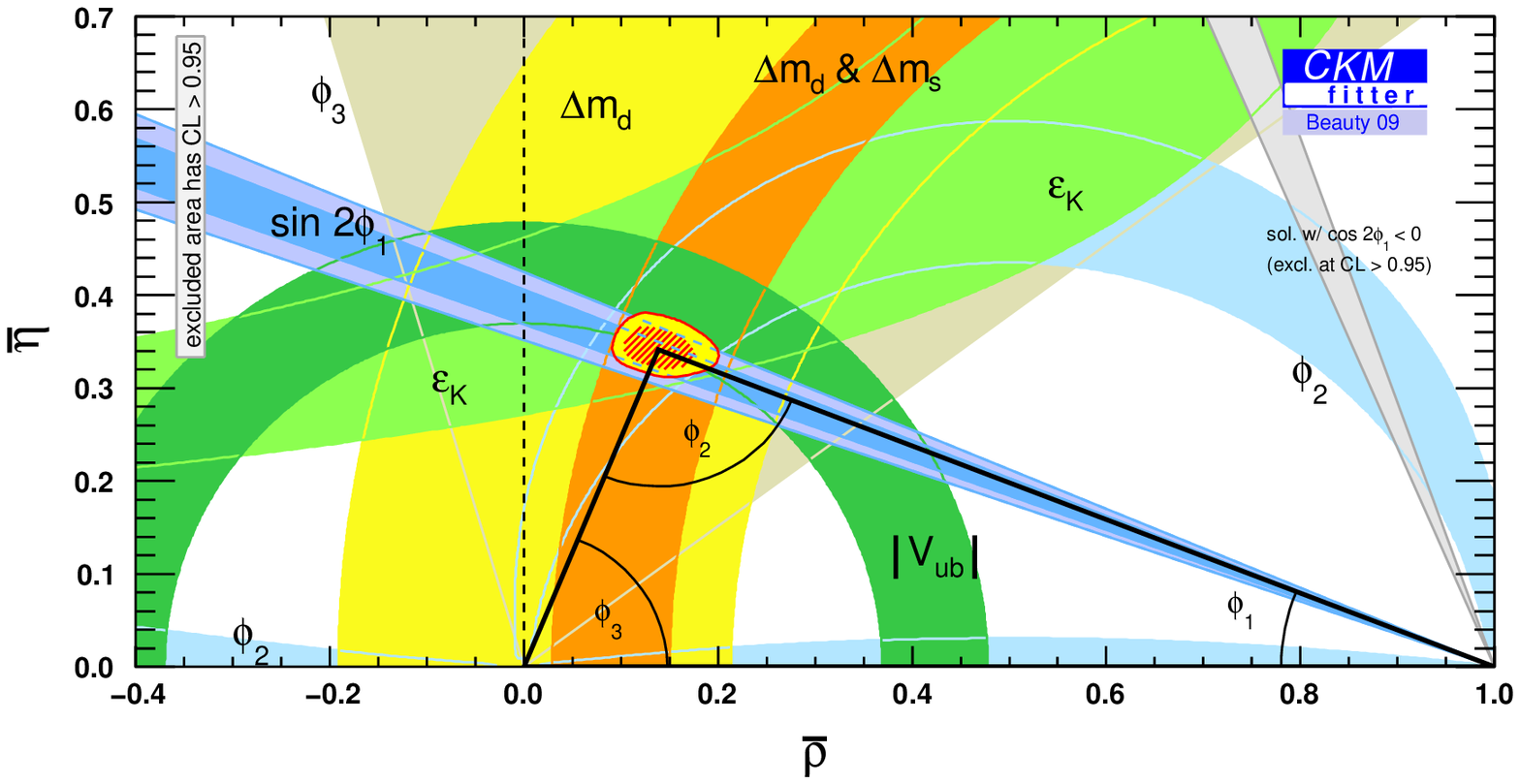}
\caption{Constraints on the UT coming from the measurements of angles
         only (left) and using all relevant experimental inputs (right).}
\label{ckmfitresult}
\end{figure}

In Fig.~\ref{ckmfitresult} we summarize the constraints on the UT coming
from the measurements of angles only, as well as after including other
experimental inputs. To a very good approximation, the Kobayashi-Maskawa
formalism is found to be the right description of \CP\ violation in the SM.
Needless to say that we still need to improve the precision on the third
angle $\phi_3$ -- one has just made a head-start! Similarly, we expect
the errors on other two angles to shrink further, {\it e.g.}, once Belle
analyzes its full \FourS\ dataset.

\section{Search for physics beyond the SM}

In this section we attempt to enumerate various hints for, or constraints
on, potential new physics contributions, as observed with the $B$ factories.

\subsection{Measured {\boldmath $\sin(2\phi_1)$} with the penguins}

\begin{figure}[!htb]
\center
\includegraphics[width=.7\columnwidth]{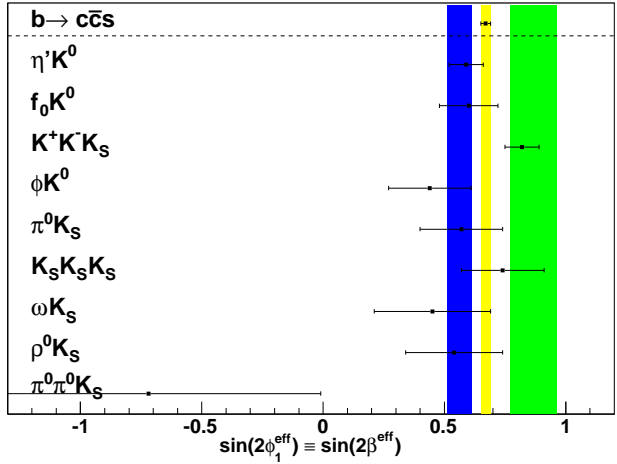}
\caption{Measured values of $\sin(2\phi_1)$ in (yellow/light-shaded)
   charmonium decays, (blue/dark-shaded) penguin decays, and
   (green/medium-shaded) inferred from indirect measurements~\cite{lunghisoni}.}
\label{adrian}
\end{figure}

As $\sin(2\phi_1)$ is the most precisely measured observable concerning
\CP\ violation in $B$ decays, we can use it as a ``Standard Candle'' to
set constraints on new physics by looking for possible deviations from
this value in a number of ways. One such is the comparison of the values
of $\sin(2\phi^{\rm eff}_1)$ measured in penguin dominated decays with
the world-average value of $\sin(2\phi_1)$, coming from decays involving
charmonium final states. The caveat to making such a comparison is that
the penguin modes may have additional topologies that could lead to a
difference between $\sin(2\phi_1)$ and $\sin(2\phi^{\rm eff}_1)$. If these
SM corrections, $\Delta_{\rm SM}$, are well known then any residual difference
$\Delta S=\sin(2\phi^{\rm eff}_1)-\sin(2\phi_1)-\Delta_{\rm SM}$ would be
from new physics. It has been recently pointed out~\cite{lunghisoni} that
by comparing the penguin to tree channels one remains insensitive to possible
new physics contribution common to both. Therefore, it is important to compare
the directly measured values of $\sin(2\phi^{\rm eff}_1)$ with the predictions
of SM-based constraints for the same observable. Figure~\ref{adrian}
summarizes the different constraints on $\sin(2\phi^{\rm eff}_1)$, where
the maximum difference between the measured and indirect values has a
significance above $2$ standard deviations.

\subsection{Direct {\boldmath $\CP$} violation in {\boldmath $B$} decays}

Both Belle and BaBar have carried out a number of sensitive \CP\ violation
measurements in various $B$ decays. Most notable of them is the decay
$B^0\to K^+\pi^-$, where direct \CP\ violation has been established beyond
any doubt -- the measured \CP\ asymmetry is $(-9.8^{+1.2}_{-1.1})\%$.
There are a number of interesting evidences at the level of 3 standard
deviations in the decays $B^0\to\eta K^{*0}$, $B^-\to\eta K^-$,
$B^-\to\rho^0 K^-$, $B^0\to\rho^+\pi^-$ and $B^-\to\Db^{(*)0}K^-$. Another
important result has come out from $B^-\to K^-\pi^0$, with the \CP\ asymmetry
$(+5.0\pm 2.5)\%$. This in contrast to the result of $B^0\to K^+\pi^-$~\cite{nature},
where similar Feynman diagrams contribute at the tree level, tells us that
it could be either due to a large contribution from the color-suppressed
tree diagram, or from possible new physics contribution in the electroweak
penguin, or from both. Before firmly concluding anything, it is
suggested~\cite{sumrule} to check the \CP\ violation result from the decay
$B^0\to K^0\pi^0$, with a larger dataset.

\subsection{Polarization puzzle in {\boldmath $B\to VV$}}

For a $B$ meson decaying to two vector particles, $B\to VV$, theoretical
models based on QCD factorization~\cite{vvqcdf} or perturbative
QCD~\cite{vvpqcd} predict the fraction of longitudinal fraction $f_L$
to be approximately $1-(m^2_V/m^2_B)$, where $m_{V(B)}$ is the mass of
the vector ($B$) meson, for tree-dominated decays. As an example, in
the case of $B\to\rho\rho$ the prediction for $f_L$ is close to $0.9$,
which matches well with the measurement~\cite{hfag}. For
decays dominated by the penguin transition, however, there is a large
discrepancy between predictions ($\sim 0.75$) and observations, that
tend to cluster around $0.5$. This unexpected result on polarization,
mostly driven by the measurement of $B\to\phi K^*$, has motivated
several further studies.

\subsection{Constraints on the charged Higgs}

\begin{figure}[!htb]
\center
\includegraphics[width=.5\columnwidth]{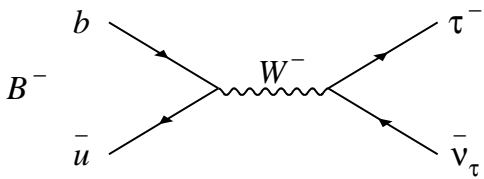}
\caption{Purely leptonic $B$ decays proceed via the annihilation of
   quark-antiquark into a $W$ boson (or, potentially into a charged
   Higgs boson).}
\label{taunu}
\end{figure}
The purely leptonic decay $B^-\to\tau^-\nutb$ provides an excellent
probe for the charged Higgs that could potentially appear in the
annihilation of $b$ and $\bar{u}$ quarks similar to the SM diagram,
where a $W^-$ boson is created in the annihilation process (see
Fig.~\ref{taunu}). For instance, if we take the prediction of the
two-Higgs doublet model~\cite{2hdm}, the observed branching fraction could
be enhanced or suppressed by a factor of $(1-m^2_B\tan^2\beta/m^2_H)^2$,
where $m_H$ is the mass of the charged Higgs and $\tan\beta$ is the
ratio of the two Higgs vacuum expectation values. On the experimental
side, identifying the decay $B^-\to\tau^-\nu_\tau$, which involves at
least two neutrinos in the final state, is a real challenge. Both
Belle and BaBar have made the best use of their detector hermiticity
and particle identification capability, and in doing so they
obtain~\cite{taunuresult} a branching fraction world-average of
$(1.73\pm 0.35)\times 10^{-4}$ for the decay. The SM prediction is
$(1.20\pm 0.25)\times 10^{-4}$, where the dominant uncertainties
come from the error in the CKM matrix element $V_{ub}$ and the
$B$-meson decay constant. Comparing the SM expectation with the
measurement, we derive a constraint on $m_H$ as a function of
$\tan\beta$. This constraint is well complimented by the measurement
of $B\to D^{(*)}\tau\nu_\tau$~\cite{dtaunuresult} and the inclusive
$b\to s\gamma$ measurement~\cite{bsgamma}. It is worth noting
that the combined result~\cite{iijima}, which excludes a charged
Higgs up to a mass of $600\gevcc$ for $\tan\beta>60$ and $300\gevcc$
for $\tan\beta>30$, is already comparable to what is expected for a
direct search~\cite{atlas} using a $30\invfb$ data sample at the LHC.

\subsection{{\boldmath $B\to K^{(*)}\ell^+\ell^-$}: Any smoking gun?}

\begin{figure}[!htb]
\center
\includegraphics[width=.7\columnwidth]{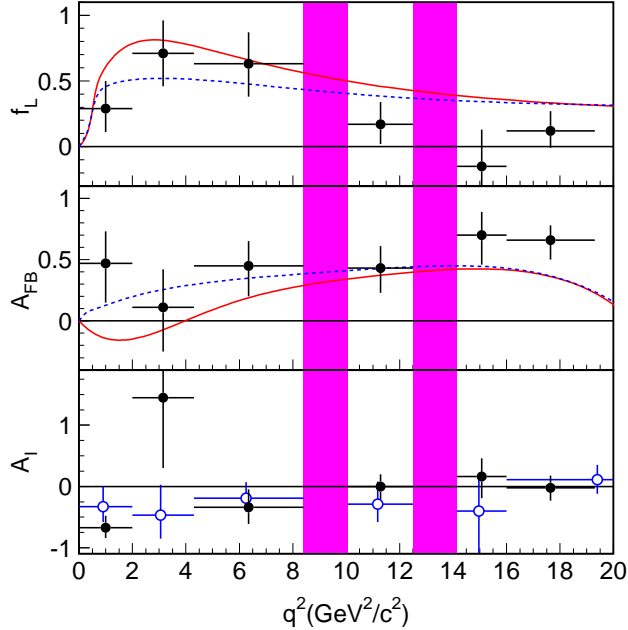}
\caption{Results for (top) $f_L$ and (middle) $A_{FB}$ in $K^*\ell^+\ell^-$
   as a function of $q^2$, together with the solid (dotted) curve representing
   the SM ($C^{\rm NP}_7=-C^{\rm SM}_7$) prediction. (Bottom) The plot of
   $A_I$ {\it vs.} $q^2$ for the $K^*\ell^+\ell^-$ (filled circles) and
   $K\ell^+\ell^-$ (open circles) modes. The two shaded regions are veto
   windows to reject events containing a $\jpsi$ or a $\psitwos$.} 
\label{afb-belle}
\end{figure}

The decay channel $b\to s\ell^+\ell^-$ is an experimenters delight, since
it offers many interesting observables that can be measured in the decays
of $B$ mesons to both inclusive and exclusive $s\ell^+\ell^-$ final
states, where $s$ denotes a strangeness-one meson. In particular, for the
exclusive mode $K^{(*)}\ell^+\ell^-$ the observables include $f_L$,
the forward-backward asymmetry $A_{FB}$, the isospin asymmetry $A_I$, and
the ratio of rates to $e^+e^-$ and $\mu^+\mu^-$ final states (lepton flavor
ratio). Recent measurements at the $B$ factories~\cite{sllbelle,sllbabar}
show that the branching fraction and the lepton flavor ratio agree with
SM expectations. However, a deviation from the SM is indicated in $A_{FB}$
(Fig.~\ref{afb-belle}), albeit with large statistical uncertainty. We
need more statistics than currently available, which would be possible
with the future experiments~\cite{superb}, to either confirm or refute
this tantalizing hint. If it is finally turned out to be real, it would
be a clean signature of new physics~\cite{wilson,amoletc}.

\section{Conclusions}

The two $B$-factory experiments have performed exceptionally well, each
producing an average over 400 high-quality journal publications within
only ten years of their inception. What we present here, is a small sampling
of their recent highlighted results. It is fair to say that the SM
continues to hold its ground in the flavor sector, though there are some
hints of new physics available, which should be investigated with more data.

\section{Acknowledgements}

I thank the organizers for their kind invitation to this interesting
conference. Thanks are also due to my Belle colleagues for their
valuable help. This work is supported in part by the Department of
Atomic Energy and the Department of Science and Technology of India.

\end{document}